\documentclass[12pt,showpacs]{revtex4}
\usepackage{amsmath}
\usepackage{bm}
\usepackage{graphicx}
\begin{document}

\title{Critical sound attenuation in a diluted Ising system}

\author{P.V.~Prudnikov\footnote{ E-mail\ :\ prudnikp@univer.omsk.su}, V.V.~Prudnikov}

\affiliation{ Dept. of Theoretical Physics, Omsk State University 55a, Pr. Mira, 644077, Omsk, Russia}

\begin{abstract}
The field-theoretic description of dynamical critical effects of the influence of
disorder on acoustic anomalies near the temperature of the second-order phase
transition is considered for three-dimensional Ising-like systems. Calculations
of the sound attenuation in pure and dilute Ising-like systems near the critical
point are presented. The dynamical scaling function for the critical attenuation
coefficient is calculated. The influence of quenched disorder on the asymptotic
behaviour of the critical ultrasonic anomalies is discussed.
\end{abstract}


\maketitle

\begin{figure}
\includegraphics[width=0.6\textwidth]{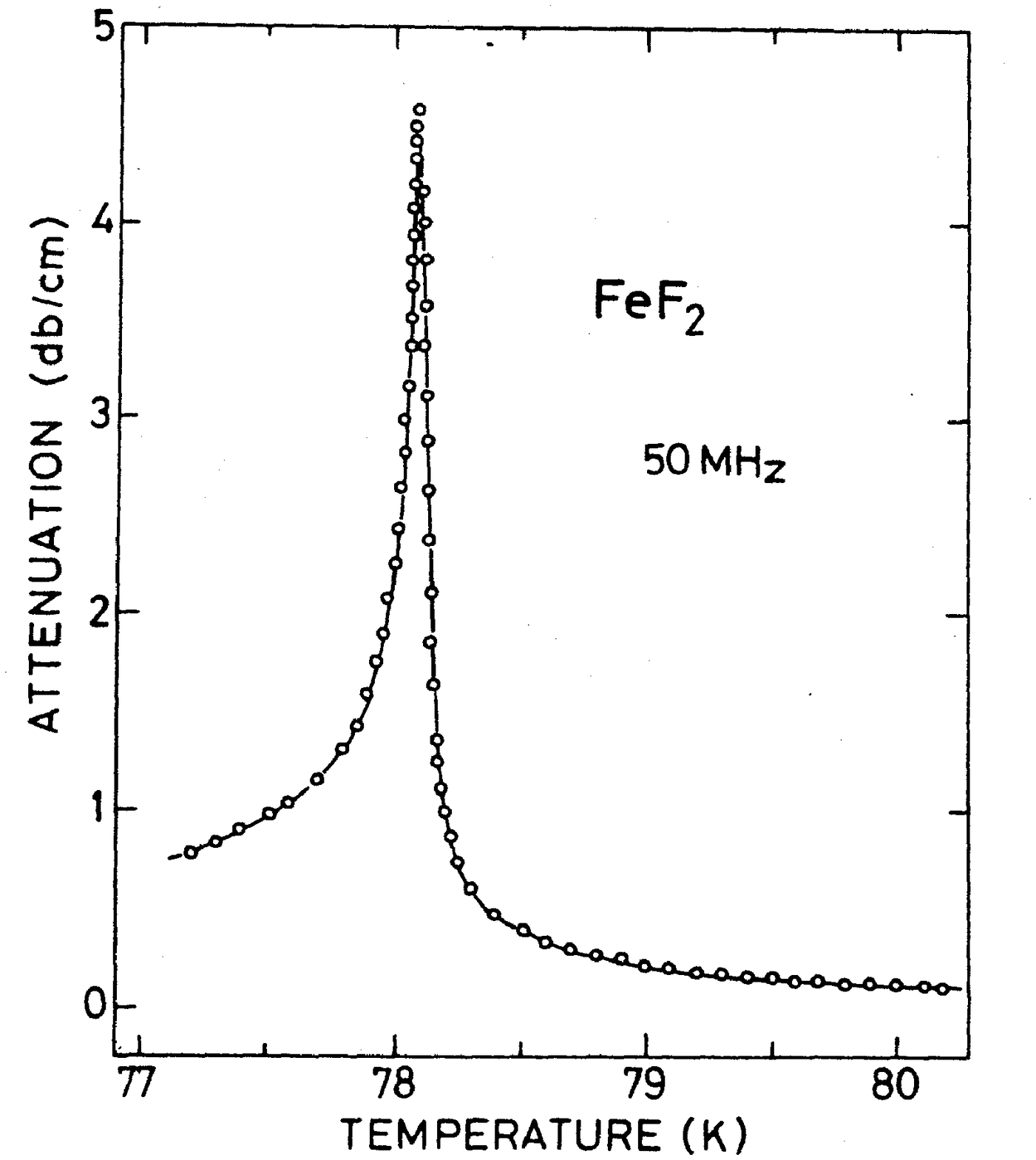}
\caption{ \label{fig:1} Ultrasonic anomaly for experimental attenuation in $\mathop{\mathrm{Fe}}\mathrm{F}_2$ \protect\cite{IkushimaF}. }
\end{figure}

The progress achieved in the understanding of critical phenomena has largely been due to
theoretical and experimental works devoted to studying the critical dynamics of condensed
media. We have seen from experiments (Fig.~\ref{fig:1}) \cite{IkushimaF} that for a solid an anomalous peak
of ultrasonic attenuation is observable in the vicinity of the critical point. The critical
anomalies exhibited by sound attenuation have long been recognized as important in the study
of dynamical critical phenomena. Ultrasonic methods permit simultaneous measurements of
both static and dynamic properties. Measurements of sound velocities give information on the
equilibrium properties, while measurements of the sound attenuation yield information on the
dynamic properties ofmaterials. The main difficulty in the theoretical discussion of the critical
propagation of sound waves consists in the estimation of the four-spin correlation function.
The method which is based on the representation of the four-spin correlation function through
two-spin correlation functions by means of decoupling leads to overestimated values of the
critical fluctuations.

There are a lot of theories and phenomenological descriptions
\cite{LandauKh,Kawasaki,Matching,IroSchwabl,Pawlak,Schwabl93,Kamilov89,Aliev89} of ultrasonic anomalies in solids with good agreement 
with experiments \cite{Bhatt,Luthi,Suzuki82} but real materials and crystals have many 
structural defects and it is worth considering the influence of such defects or disorder 
on the dynamical process of sound propagation in solid media.
Structural disorder and the presence of impurities or other defects play an important role in 
real materials and physical systems. They may induce new phase transition types of and 
universality classes and modify the dynamic transport properties.

According to the Harris criterion \cite{Harris74}, critical behaviour of Ising systems is 
changed by presence of a weak quenched disorder.
The problem of the influence of disorder on the critical sound propagation in Ising-like 
systems has been discussed in \cite{PawlakFecher89} with use of a $\varepsilon$-expansion 
in the lowest order of approximation. 
However, our pilot analysis of this phenomenon showed that in \cite{PawlakFecher89} 
some diagrams which are needed for a correct description of 
the influence of the disorder were not considered.
Furthermore, our numerous investigations of pure and disordered systems
performed in the two-loop and higher orders of the approximation
for the three-dimensional system directly, together with the use of methods of series 
summation, show that the predictions made in the lowest order of the approximation,
especially on the basis of the $\varepsilon$ - expansion, can differ
strongly from the real critical behaviour \cite{Prudnikov01_03,HighOrder}. Therefore, the results
from \cite{PawlakFecher89} must be reconsidered with the use of a more accurate 
field-theoretic approach in the higher orders of the approximation.

In this paper we have realized the correct field-theoretic description of dynamical effects of
the influence disorder on acoustic anomalies near the temperature of the second-order phase 
transition for the three-dimensional Ising-like systems in the two-loop approximation.

In our description, we extended the model of the phase transition in a disordered system 
with a coupling between nonfluctuating variables \cite{Skryabin} to the case 
(physically important for structural phase transitions) of a compressible 
three-dimensional Ising model with frozen-in lattice defects that is considered 
in using the renormalization-group method in the two-loop approximation. 

The interaction of the order parameter with elastic deformations plays a significant role in 
the critical behaviour of the compressible system. It was shown for the first time in \cite{Larkin69} 
that the critical behaviour of a system with elastic degrees of freedom is 
unstable with respect to the connection of the order parameter with acoustic modes and   
a first-order phase transition is realized. However, the conclusions of \cite{Larkin69}
are only valid at low pressures. 
It was shown in \cite{Ymry74} that in the range of high pressures, beginning from a threshold 
value of pressure, the deformational effects induced by the external pressure lead to a 
change in type of the phase transition.

The Hamiltonian of a disordered Ising model with allowance for elastic degrees of freedom 
may be specified as
\begin{equation}
\label{ham:1} H = H_{el} + H_{op} + H_{int} + H_{imp},
\end{equation}
consisting of four contributions.

The elastic part is determined from
\begin{equation}
H_{el}=\frac{1}{2}\int {\rm d^{d}}x\, \left[C^{0}_{11}\sum\limits_{\alpha} u^{2}_{\alpha \alpha}
 + 2C^{0}_{12}\sum\limits_{\alpha\beta} u_{\alpha \alpha} u_{\beta \beta}
 + 4C^{0}_{44} \sum\limits_{\alpha <\beta}u^{2}_{\alpha\beta}\right],
\end{equation}
where $u_{\alpha \beta}(x)$ are components of the strain tensor
and $C_{ij}^{k}$ are the elastic moduli.

$H_{op}$ is a magnetic part in the appropriate Ginzburg-Landau form:
\begin{equation}
H_{op} = \int  {\rm d^{d}}x\, \left[ \frac{1}{2}\tau_0 S^{2} + \frac{1}{2}\left(\nabla S\right)^{2} +
\frac{1}{4}u_0 S^{4} \right],
\end{equation}
where $S(x)$ is the Ising field variable which is associated with the spin order parameter,
$u_0$ is a positive constant and $\tau_0 \sim (T - T_{0c})\left/ T_{0c}\right.$ with the mean-field 
phase transition temperature $T_{0c}$. 

The term $H_{int}$ describes the spin-elastic interaction
\begin{equation}
H_{int} = \int {\rm d^{d}}x\, \left[ g_{0} \sum_{ \alpha} u_{\alpha \alpha} S^{2} \right],
\end{equation}
which is bilinear in the spin order parameter and linear in deformations.
$g_{0}$ is the bare coupling constant.

The term $H_{imp}$ of Hamiltonian determines the influence of disorder and
it is considered in the following form:
\begin{equation}
H_{imp} = \int {\rm d^{d}}x\, \left[\Delta{\tau}(x)S^{2}\right]+\int{\rm d^{d}}x\, \left[h(x)\sum\limits_{\alpha} u_{\alpha\alpha}\right],
\end{equation}
where the random Gaussian variables $\Delta{\tau}(x)$ and $h(x)$ are the local 
transition temperature fluctuations and induced random stress, respectively.
Taking the $\Delta{\tau}(x)$ fluctuations into account causes additional 
interaction of order parameter fluctuations over defects and renormalization 
of the phase transition temperature for disordered systems. Taking into account 
the $h(x)$ fluctuations leads to renormalization of the elastic moduli 
and the coupling constant in the spin-elastic interaction.

The Fourier transformed variables become
\begin{equation}
\label{eq:Ftr}
 u_{\alpha\beta} = u^{(0)}_{\alpha \beta} + V^{-1/2} \sum_{q \neq 0}
 u_{\alpha\beta}(q) \exp\left(i q x\right),
\end{equation}
with $u_{\alpha \beta}(q) = {\rm i}/2\left[q_\alpha u_\beta+q_\beta u_\alpha\right]$. 
Then the normal-mode expansion is introduced as 
$\vec{u}(q)=\sum_\lambda \vec{e}_\lambda(q) Q_{q,\lambda}$ with the normal coordinate 
$Q_{q,\lambda}$ and polarization vector $\vec{e}_\lambda(q)$.
We carry out the integration in the partition function with
respect to the nondiagonal components of the uniform part of the deformation tensor
$u^{(0)}_{\alpha\beta}$, which are insignificant for the critical behaviour of the
system in an elastically isotropic medium.

After all of the transformations \cite{Prudnikov01} the effective Hamiltonian of the system has become 
\begin{equation}
\label{ham:8}
\begin{split}
\tilde{H} =& \frac{1}{2}\int {\rm d^{d}}q\ \left(\tau_0+q^{2}\right)\, S_{q}\, S_{-q} 
	  + \frac{1}{2}\int {\rm d^{d}}q\ \Delta{\tau}_{- q}\, S_{q_1}\, S_{q-q_1}  \medskip \\
	  &+ \frac{1}{4}u_0 \int {\rm d^{d}}q\ S_{q_1}\, S_{q_2}\, S_{q_3}\, S_{-q_1-q_2-q_3}
	  + \int {\rm d^{d}}q \ q\,h_{-q}\, Q_{q, \lambda} \medskip \\
	  &- \frac{1}{2}w_0 \int {\rm d^{d}}q\ \left(S_{q}\, S_{-q}\right)\, \left(S_{q}\, S_{-q}\right)
	  -g_0 \int {\rm d^{d}}q\ q\, Q_{-q, \lambda}\, S_{q_1}\, S_{q-q_1} \medskip \\
	  &+ a_0 \int {\rm d^{d}}q\ q^{2}\, Q_{q, \lambda}\, Q_{-q, \lambda},
\end{split}
\end{equation}
where $w_0=3g_0^2\left/ \left( 2V\left( 4C_{12}^0 - C_{11}^0 \right)\right)\right.$;
$a_0=(C_{11}^0+4C_{12}^0-4C_{44})\left/ 4V \right.$ 

The renormalization-group analysis of the critical behaviour of the disordered compressed 
Ising model with the Hamiltonian (\ref{ham:8}) was carried out in our paper \cite{Prudnikov01}, 
and different fixed points for the Hamiltonian (\ref{ham:8}) and conditions of their stability 
were determined in the two-loop order of the approximation with the use of the Pad\'{e}-Borel 
summation technique. It was shown that the $\Delta{\tau}_{q}$ fluctuations are relevant 
for the critical behaviour of the Ising model and the replica averaging procedure over 
$\Delta{\tau}_{q}$ leads to a new vertex of interaction for the order parameter fluctuations. 
It was shown in \cite{Prudnikov01} that random stress connected with $h_{q}$ in (\ref{ham:8})  
can lead to multicritical behaviour of a system if certain conditions are fulfilled.

The critical dynamics of the system in the relaxational regime can be described by the 
Langevin equations \cite{FirstLangevin} for the spin order 
parameter $S\left( q\right)$ and phonon normal coordinates $Q_{\lambda}\left( q\right)$
\begin{eqnarray}
\label{dinamic:1}
\dot{S}_{q}=-\Gamma_{0}\,\frac{\partial\tilde{H}}{\partial S_{-q}}+\xi_{q}+\Gamma_{0} h_{S}, \nonumber \\
\label{dinamic:2} 
\ddot{Q}_{q, \lambda}=-\frac{\partial\tilde{H}}{\partial Q_{-q, \lambda}} -
q^{2} D_{0} \dot{Q}_{q, \lambda}+\eta_{q, \lambda}+h_{Q},
\end{eqnarray}
where $\Gamma_{0}$ and $D_{0}$ are the initial kinetic coefficients, 
$\xi_{q}(x,t)$ and $\eta_{q}(x,t)$ are Gaussian white noises.

The quantities of interest are the response functions $G(q,\omega)$ and $D(q,\omega)$ 
of spin and deformation  variables, respectively. It can be obtained by linearization in correspondent fields that 
\begin{eqnarray}
D(q,\omega) &=& \delta \left[\langle{Q_{q,\omega}}\rangle\right]\left/\delta{h_{Q}}\right. 
= \left[\langle Q_{q,\omega}Q_{-q,-\omega}\rangle\right], \medskip \\
G(q,\omega) &=& \delta \left[\langle{S_{q,\omega}}\rangle\right]\left/\delta{h_{S}}\right. 
= \left[\langle S_{q,\omega}S_{-q,-\omega}\rangle\right],
\end{eqnarray}
where $\langle ... \rangle$ denotes averaging over Gaussian white noises, 
$\left[ ... \right]$ denotes averaging over random fields $\Delta{r}_{q}$ and $h_{-q}$.

The response functions may be expressed in terms of self-energy parts: 
\begin{eqnarray}
\label{Dys1}
G^{-1}(q,\omega) &=& G_{0}^{-1}(q,\omega) + \Pi(q,\omega), \\
D^{-1}(q,\omega) &=& D_{0}^{-1}(q,\omega) + \Sigma(q,\omega), \nonumber
\end{eqnarray}
where the free response functions $G_0(q,\omega)$ and $D_0(q,\omega)$ have the forms
\begin{eqnarray}
D_{0}(q,\omega) &=& 1\left/\left(\omega^{2}-a q^{2}-\mathrm{i}\omega D_{0} q^{2}\right)\right., \nonumber\\
G_{0}(q,\omega) &=& 1\left/\left(\mathrm{i}\omega\left/\Gamma_{0}\right.+\left(\tau_0+q^{2}\right)\right)\right.. \nonumber
\end{eqnarray}

The characteristics of the critical sound propagation are defined by means of the response function $D(q,\omega)$. 
Thus, the coefficient of ultrasonic attenuation is determined through 
imaginary part of $\Sigma(q,\omega)$: 
\begin{equation}
\label{Atten}
    \alpha(\omega,\tau) \sim \omega\mathop{\mathrm{Im}}\Sigma(0, \omega).
\end{equation}

\begin{figure}
\includegraphics[width=0.7\textwidth,keepaspectratio]{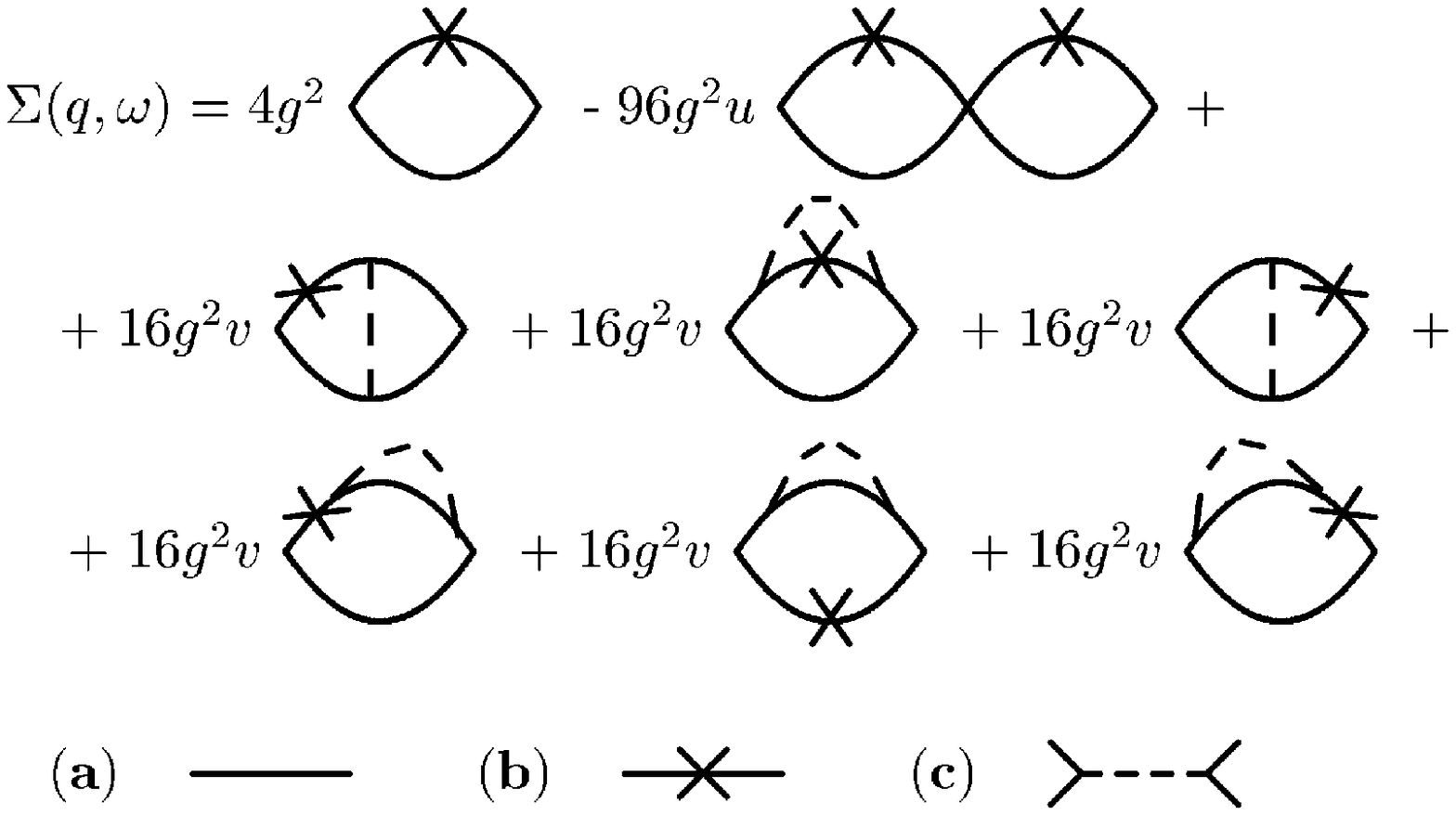}
\caption{ \label{fig:2} The diagrammatic representation of the $\Sigma(q,\omega)$ in the two-loop approximation. 
Solid line ($\mathbf{a}$) corresponds to $G_{0}(q,\omega)$, solid line with cross ($\mathbf{b}$) is $2\,G_{0}(q,\omega)G_{0}(-q,-\omega)\left/ \Gamma_0 \right.$, 
and vertex with dash line ($\mathbf{c}$) corresponds to  $v=\bigl[(\Delta \tau)^2\bigr]$ }
\end{figure}

The self-energy part $\Sigma(q,\omega)$ was obtained by iterative solution \cite{Ma} 
of the dynamic equation (\ref{dinamic:1}) with the effective Hamiltonian (\ref{ham:8}).
The diagrammatic representation of $\Sigma(q,\omega)$ in the two-loop approximation is presented in Fig.~\ref{fig:2}.

The Feynman diagrams involve momentum integrations on dimension $d$ (in our
case $d=3$). Near the critical point the correlation length $\xi$ increases
infinitely. When $\xi^{-1} \ll \Lambda$, where $\Lambda$ is a cutoff in
momentum-space integrals (the cutoff $\Lambda$ serves to specify the basic
length scale), the vertex functions are expected to display an asymptotic
scaling behaviour for wave-numbers $q \ll \Lambda$. Therefore, one is lead to
consider the vertex functions in the limit $\Lambda \rightarrow \infty$. 
The use of the renormalization-group scheme removes all divergences which arise 
in thermodynamic variables and kinetic coefficients in this limit.

We have applied the 
matching method which was introduced for statics in \cite{Nelson76} and generalized 
for critical dynamics in \cite{Matching}.
First, we use the dynamical scaling property of the response function
\begin{equation}
\label{RespF}
D\left(q,\omega,\tau\right) = e^{(2-\eta)l} D\left(q e^{l},\left(\omega/\Gamma_0\right) e^{zl},\tau e^{l/\nu}\right),
\end{equation}
and then we calculate the right-hand side of this equation for some value $l=l^*$,
where the arguments do not all vanish simultaneously \cite{Nelson76}.
The choice of $l^*$ is determined by
\begin{equation}
\label{MCond}
\left[\left(\omega\left/\Gamma_{0}\right. \right) e^{z l^*}\right]^{4/z} + 
\left[\left( \tau e^{l^*/\nu} \right)^{2\nu}+q^2 e^{2l^*}\right]^2=1.
\end{equation}

It may be argued \cite{Matching} that condition (\ref{MCond}) provides an infrared cut-off 
for all divergent values. The particular form of the matching condition (\ref{MCond})
containing the exponents $z$ and $\nu$ permits an explicit solution for $l^*$
\begin{equation}
\label{el}
e^{l^*}=\tau^{-\nu} \left[ \,1 +\left(y/2\right)^{4/z} \right]^{-1/4}\equiv \tau^{-\nu}F(y,\tau) .
\end{equation}

In (\ref{el}) the abbreviation $y=\omega\tau^{-z \nu}\left/\Gamma_{0}\right.$ 
is introduced and $F(y,\tau)$ is defined. The value of the exponent $\nu=0.70$ 
was calculated in \cite{Prudnikov01} for the corresponding fixed point. The exponent $z=2.1653$ was taken from \cite{Jetp98} 
for the 3D disordered Ising model considered for purely relaxational model $A$. 
The coupling of the order parameter with elastic deformations is irrelevant 
for the dynamics of a disordered system with a negative specific heat exponent \cite{HoenHalp}.

The response function $D\left(q e^{l},\left(\omega/\Gamma_0\right) e^{zl},\tau e^{l/\nu}\right)$ on the right hand side of (\ref{RespF}) 
is represented by the Dyson equation (\ref{Dys1}) and for the self-energy part we obtain
\begin{eqnarray}
\label{ImSelfR}
\displaystyle\frac{ {\rm Im}\Sigma(\omega)}{\omega} = 
\exp\left(l\frac{\left(\alpha+z\nu\right)}{\nu}\right)
\displaystyle\frac{ {\rm Im}\Sigma(\omega e^{zl})}{\omega e^{zl}}.
\end{eqnarray}

It was shown in later theoretical works \cite{Kawasaki,Suzuki82} that in asymptotic regions 
the coefficient of attenuation can be described using a simple scaling function of the variable  
$y$ only. The experimental investigations performed on the three-dimensional 
crystals $\mathop{\mathrm{Gd}}$ \cite{Luthi} and $\mathop{\mathrm{Mn}\mathrm{P}}$ \cite{Suzuki82} confirmed the validity of
the concepts of dynamical scaling.

Thus after renormalization procedure (\ref{ImSelfR}) we can define a scaling relation in the form 
\begin{equation}
\label{norm:20}
\mbox{Im}\Sigma(\omega)\left/\omega\right. = \tau^{-\alpha-z\nu}\phi(y), 
\end{equation} 
where $\phi(y)$ is a dynamical scaling function
\begin{equation} \label{ScaleF}
\begin{split}
\phi(y) = \displaystyle\frac{g^{*2}\Gamma_0}{\pi\mathstrut} &
	\displaystyle\frac{F^{\alpha/\nu+1/2\nu-z}}{y^2}
\left\{ 1- \left[ \displaystyle\frac{1}{2}\left(1+\displaystyle\frac{y^2F^{2z-2/\nu}}{4}\right)^{1/2}
        +\displaystyle\frac{1}{2}\right]^{1/2}  \right\} \bigskip \\
 &-\displaystyle\frac{ 12g^{*2}u^*\Gamma_0^2}{\pi^2\mathstrut}
	\displaystyle\frac{ F^{\alpha/\nu+1/\nu-2 z}}{ y^3}
        \left\{ \left[\displaystyle\frac{1}{2}\left(1+\displaystyle\frac{y^2F^{2z-2/\nu}}{4}\right)^{1/2}-\displaystyle\frac{1}{2}\right]^{1/2}
        -\displaystyle\frac{yF^{z-1/\nu}}{4} \right\} \bigskip  \\
 &+\displaystyle\frac{8g^{*2}v^*}{(4\pi)^3}
        \displaystyle\frac{F^{\alpha/\nu-z}}{y^2}
        \,\phi_{\rm imp}(y), 
\end{split}
\end{equation}
where $g^*$, $u^*$, and $v^*$ are values of vertices in the fixed point of renormalization group transformations \cite{Prudnikov01},
$\phi_{\rm imp}\left(y\right)$ is the numerically calculated contribution of diagrams in $\Sigma\left(\omega\right)$ (Fig.~\ref{fig:2}) characterizing 
the influence of disorder. 

\begin{figure}
\includegraphics[width=0.8\textwidth,keepaspectratio]{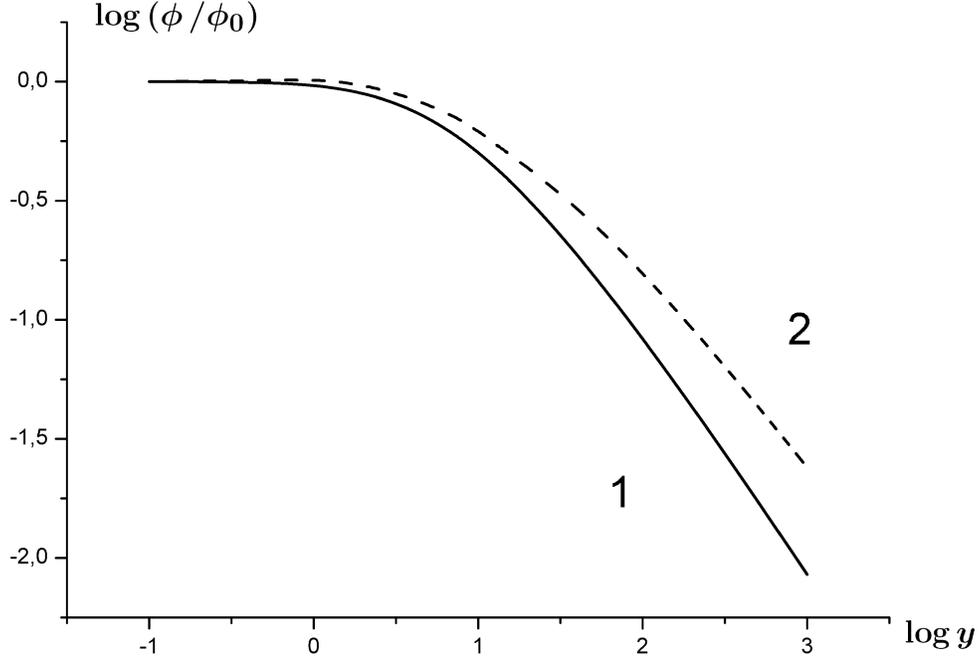}
\caption{ \label{fig:3} Scaling functions for the critical sound attenuation $\phi(y)$ in double 
logarithmic plotting for pure (1) and disordered (2) Ising systems ($\phi_0=\phi(0)$).  }
\end{figure}

The dynamical scaling function $\phi(y)$ is plotted against $y$ on a double-logarithmic scale 
for pure and disordered systems in Fig.~\ref{fig:3}.
We have thus seen that the presence of disorder is irrelevant for the scaling behaviour in the 
hydrodynamical region with $y\ll 1$, but it has a drastic effect in the 
critical region with $y\gg 1$ ($T\rightarrow T_c$). 

\begin{figure}
\includegraphics[width=0.6\textwidth]{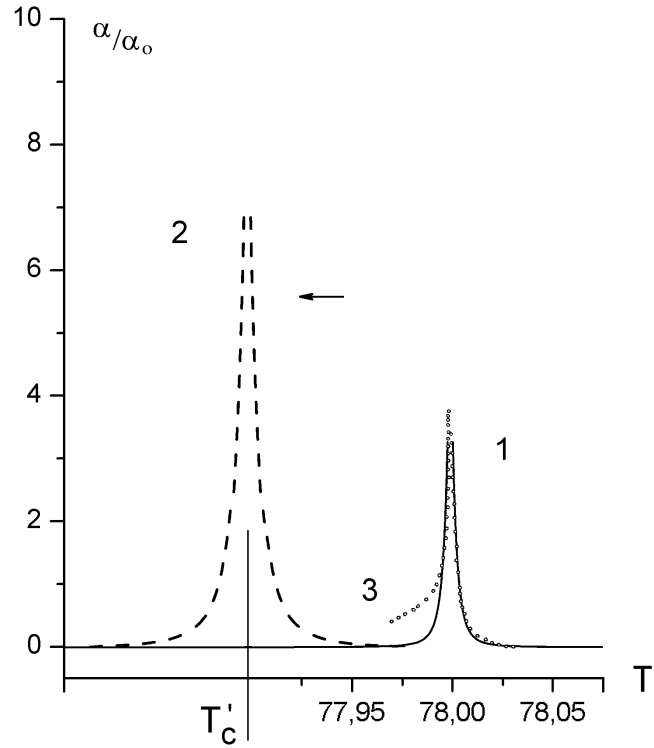}
\caption{ \label{fig:4} Thermal dependencies of the ultrasonic attenuation in critical point for pure (1) 
and disordered (2) systems in comparison with experimental results (3) in $\mathop{\mathrm{Fe}}\!\mathrm{F}_2$ \protect\cite{IkushimaF}.}
\end{figure}

The anomalous temperature dependence of the calculated attenuation coefficient for pure and disordered systems is shown in Fig.~\ref{fig:4}. 
Comparison of curves (1) and (2) in Fig.~\ref{fig:4} clearly reveals the strong influence of disorder on the 
temperature dependence of the attenuation coefficient in the vicinity of the critical point. 
We hope that these theoretical results will create a demand for ultrasonic experimental 
investigations of dilute Ising-like systems, for example samples 
of $\mathop{\mathrm{Fe}}_x\!\mathop{\mathrm{Zn}}_{1-x}\!\mathrm{F}_2$. 
In producing the figure we used the fact that the presence of disorder causes a reduction of the phase transition 
temperature $T_c$ in relation to that of the pure system. 
A model representation of the calculated attenuation coefficient for the pure system is shown in Fig.~\ref{fig:4} 
in comparison with experimental data (3) for $\mathop{\mathrm{Fe}}\!\mathrm{F}_2$ \cite{IkushimaF}. 
Adjustment of the experimental data permited us to determine the value of the theoretical parameter $\Gamma_0$ and 
then calculate the attenuation coefficient for the disorder model. 
We should note that the observable differences from the experimental results below $T_c$ are 
explained by the contribution of the order parameter relaxation effects to the attenuation, which always occur below $T_c$. 
The relaxation effects are not considered in this paper, but the contribution of fluctuations to the attenuation coefficient
is relevant over the whole critical range.

From (\ref{Atten}) we find that the sound attenuation coefficient obeys the asymptotic scaling relation

\begin{equation}
\label{ass:1}
\alpha\left(\omega,\tau\right)\sim\omega^{2}\tau^{-\alpha-\nu{z}}\phi\left(y\right).
\end{equation}

\begin{table}[t]
\begin{center}   
\caption{\label{tab:1} Asymptotical behaviour of the attenuation coefficient in the hydrodynamical and the critical regions}
\begin{tabular}{lll} \hline\hline 
 System          & Region                 & $\alpha(\omega, \tau)$          \\ \hline 
 Pure            & hydrodyn.              & $\omega^{2   }\tau^{-1.37}$  \\
                 & critical               & $\omega^{1.04}\tau^{-0.16}$  \\            \hline
 Disordered      & hydrodyn.              & $\omega^{2   }\tau^{-1.44}$  \\ 
                 & critical               & $\omega^{1.23}\tau^{-0.26}$  \\ \hline\hline 
\end{tabular}
\end{center}
\end{table}

On the basis of the dependencies (\ref{ScaleF}) obtained, we calculated the exponents of asymptotic 
behaviour of the attenuation coefficient in the hydrodynamical and the critical regions 
(Table~\ref{tab:1}). We see from the table that in the critical region the anomalies of the attenuation coefficient 
must be observed in both pure and disordered systems. But for disordered systems the 
anomalies of the temperature and frequency dependences of the attenuation coefficient must be stronger 
than those for pure systems. These conclusions are also shown in Fig.~\ref{fig:4}.

We hope that theoretical estimates of the attenuation coefficient carried out in this paper creat a great demand for detection of the anomalies
which are found to be induced by quenched disorder in ultrasonic experimental investigations of the critical dynamics.

\begin{acknowledgments}
This work was supported by the Russian Foundation for Basic Research through Grants (No.~04-02-17524, No.~04-02-39000 and No.~05-02-16188).
\end{acknowledgments}

\end{document}